\documentclass[reprint,superscriptaddress,aps,twocolumn,showpacs]{revtex4-2}
\usepackage[centertableaux]{ytableau}
\usepackage{amssymb}
\usepackage{amsmath}
\usepackage{graphicx}
\usepackage{multirow}
\usepackage{hyperref}
\usepackage{longtable}
\usepackage{lipsum}
\usepackage{float}
\usepackage{makecell}
\usepackage{braket}

\allowdisplaybreaks

\begin{document}

\title{$P_c$ Resonances in the Compact Pentaquark Picture}

\author{W. Ruangyoo}
\email[]{wiriyahz@gmail.com}
\author{K. Phumphan}
\author{C. C. Chen}
\affiliation{School of Physics and Center of Excellence in High Energy Physics and Astrophysics, Suranaree University of Technology, Nakhon Ratchasima 30000, Thailand}
\affiliation{Department of Physics, National Cheng Kung University, Tainan, 70101, Taiwan}
\author{A. Limphirat}
\email[]{ayut@g.sut.ac.th}
\author{Y. Yan}
\email[]{yupeng@g.sut.ac.th}
\affiliation{School of Physics and Center of Excellence in High Energy Physics and Astrophysics, Suranaree University of Technology, Nakhon Ratchasima 30000, Thailand}

\date{\today}

\begin{abstract}
\indent  $P_c$ resonances are studied in the approach of quark model and group theory. It is found that there are totally 17 possible pentaquark states with the quark contents $q^3Q \bar Q$ ($q$ are $u$ and $d$ quarks; $Q$ is $c$ quark) in the compact pentaquark picture, where the hidden heavy pentaquark states may take the color singlet-singlet ($[111]_{{qqq}}\otimes [111]_{{c \bar c}}$) and color octet-octet ($[21]_{{qqq}}\otimes [21]_{{c \bar c}}$) configurations. The partial decay widths of hidden heavy pentaquark states are calculated for all possible decay channels. The results show that the $pJ/\psi$ is the dominant decay channel for both the spin $3/2$ and $1/2$ pentaquark states, and indicate that the $P_c(4440)$ may not be a compact pentaquark state while $P_c(4312)$ and $P_c(4457)$ could be the spin-$\frac{1}{2}$ and spin-$\frac{3}{2}$ pentaquark states, respectively.

\end{abstract}

\maketitle

\section{Introduction}\label{sec:Int}
\indent

In 2015, the two pentaquark-like states, $P_c(4380)$ and $P_c(4450)$, were observed for the first time at 7 and 8 TeV proton-proton collisions by the LHCb Collaboration \cite{LHCb1,LHCb2}. These two $P_c^+$ states with the minimal quark content $uudc\bar c$ lie in the $p J/\psi$ invariant mass spectra in $\Lambda_b^{0}\rightarrow p J/\psi K^{-}$ decay process. Then in 2019, the LHCb Collaboration announced the updated data of three narrow pentaquark-like states, $P_c(4312)$, $P_c(4440)$ and $P_c(4457)$ \cite{LHCb3}. In this discovery a new structure of  $P_c(4312)$ was observed in a lowest invariant mass around 4312 MeV with a width of $9.8 \pm 2.7^{+3.7}_{-4.5}$ MeV. The $P_c(4450)$ pentaquark-like state in the previous report is actually the result of two overlapped peaks--the $P_c(4440)$ and $P_c(4457)$. The two peaks have the widths of $20.6 \pm 4.9^{+8.7}_{-10.1}$ MeV and $6.4 \pm 2.0^{+5.7}_{-1.9}$ MeV, respectively. The quantum numbers of these three narrow $P_c$ states have not been confirmed by LHCb yet while $J^P=\frac{1}{2}^-$, $J^P=\frac{1}{2}^-$, and $J^P=\frac{3}{2}^-$ are only suggested for the $P_c(4312)$, $P_c(4440)$ and $P_c(4457)$ states, respectively. Moreover, the LHCb has proposed
a structure of the bound state of a baryon and a meson for the $P_c$ states stemming from that the masses of $P_c(4312)$ and $P_c(4457)$ states are approximately 5 MeV and 2 MeV below the $\Sigma_c^+ \bar{D}^0$ and $\Sigma_c^+ \bar{D}^{*0}$ thresholds, respectively, while $P_c(4440)^+$ is close to $\Sigma_c^+ \bar{D}^{*}$ threshold with a 20 MeV gap. At this moment, no further conclusion for $P_c(4380)$ state has been arrived yet.

The molecular picture of $P_c$ states have been studied theoretically in Refs. \cite{Zhu2019,Chen2019,Meng2019,Guo2019,Xiao12019,Voloshin2019,Sakai2019,Lin2019,Thomas2019} resulting from the closeness of their masses to the $\Sigma_c^+ \bar{D}^{(*)0}$ threshold and the only observed $p J/\psi$ channel. The $P_c^+$ states are also studied as compact pentaquarks  \cite{Ali2016,Ali2019,Gian2019,Weng2019}. Various approaches have been applied in theoretical studies, such as QCD sum rules \cite{Chen_2015, Chen:2019bip, Azizi_2017, Azizi_2018, Azizi_2018dva, Wang:2019got}, simple chromomagnetic model \cite{Cheng:2020irt} and others \cite{MZ_Liu:2019}. However, the nature of $P_c$ is still an open question.
	
We study in this work the $P_c$ resonances in the compact pentaquark picture  using the approach of quark model and group theory. 	
The paper is organized as follows. In Sec. \ref{sec2} all possible quark configurations and wave functions of ground state hidden-charm pentaquarks are worked out by applying the $S_3$ permutation group.  Sec. \ref{sec3} shows the calculation of the partial decay widths of the hidden-charm pentaquarks for all possible decay channels. Discussion is given in Sec. \ref{sec4}.

\section{$P_c$ wave functions in compact pentaquark picture}\label{sec2}
\indent The construction of the $P_c$ pentaquark states follows the rules that the color wave function of $P_c$ must be singlet, and the total $P_c$ wave functions should be antisymmetric under any permutation of the three light quarks. The color wave function of the $P_c$ states is required to be a ${[222]_1}$ singlet, and the total wave function for the ground $P_c$ pentaquark states is
\begin{equation}
\label{pentaquarkwave}
\Psi(q^3c \bar c)=\psi^O_{sym}\psi^C_{[222]} \psi^{SF},
\end{equation}
where $\psi^O$, $\psi^C$ and $\psi^{SF}$ are the spatial, color and spin-flavor wave functions, respectively. The spatial wave function, $\psi^O$, is fully symmetric for the ground state.

The $P_c$ wave function in this work can be considered in two parts, $q^3$ and $c\bar{c}$. For $q^3$ part, the permutation symmetry of the $q^3$ part is characterized by the $S_3$ young tabloids [3], [21], [111]. The color part of $q^3$ can be both the [111] and [21] configurations while the color part of $c\bar{c}$ are also [111] and [21], so the color singlet wave functions of $P_c$ states are either in a color singlet-singlet or a color octet-octet configuration, taking the form,
	   \begin{equation}
		 \psi^C_{[222]} = \left\lbrace
			\begin{array}{l}
			\psi^C_{[111]}(q^3)\otimes\psi^C_{[111]}(c\bar c)\\
	     	\psi^C_{[21]}(q^3)\otimes\psi^C_{[21]}(c\bar c)
			\end{array}.
			\right.					
	   \end{equation}
	   
In the color octet-octet configuration, the color wave function is composed of eight possible octet states to be in the color singlet, taking the form,
\begin{equation}
\psi^C_{[222]}= \frac{1}{\sqrt{8}}\sum^8_{i=1}\psi^C_{[21]_i,q^3}\otimes\psi^C_{[21]_i,c\bar c}.
\end{equation}

The $q^3$ part is required to be fully antisymmetric. For the color singlet-singlet configuration, the general wave function of the $q^3$ part is
\begin{equation}
	\label{eq:cosf111}
	\Psi_{[A]}^{singlet}(q^3)= \psi_{[111]}^C \psi^O_{Sym} \psi_{[3]}^{SF}.
\end{equation}
The spin-flavor part of the $q^3$ cluster is in the [21] configuration for the color octet-octet configuration, and thus the $q^3$ wave function is in the form,
\begin{equation}
	\label{eq:cosf21}
	\Psi_{[A]}^{octet}(q^3)=\frac{1}{\sqrt{2}}\psi^O_{Sym}(\psi^C_{[21]_\lambda}\psi^{SF}_{[21]_\rho}-\psi^C_{[21]_\rho}\psi^{SF}_{[21]_\lambda}).
	\end{equation}
The $\lambda$ and $\rho$  color wave functions are listed in Appendix \ref{App:colorwave}. The possible spin-flavor configurations of the $q^3$ cluster, listed in Table \ref{Tab:SFw}, are taken from the previous studies in Refs. \cite{Yan:2011,Sorakrai2012,Kai2019PRC,Kai2020PRD}.
	\begin{table}[H]
 	\caption{Spin-flavor configurations of the $q^3$ cluster. \label{Tab:SFw}}
 	\def\arraystretch{1.5}
  		\begin{ruledtabular}
   		\begin{tabular}{c| c c}
    	$[3]_{SF}$ & $[3]_{F}[3]_{S}$ & $[21]_{F}[21]_{S}$\\
   		\hline
   		$[21]_{SF}$ & \makecell{$[3]_{F}[21]_{S}$ \\ $[21]_{F}[3]_{S}$} & \makecell{$[21]_{F}[21]_{S}$ \\ $[111]_{F}$ $[21]_{S}$}\\
   		\hline
   		$[111]_{SF}$ & $[21]_{F}[21]_{S}$ & $[111]_{F}[3]_{S}$ \\
   		\end{tabular}
   		\end{ruledtabular}
 	\end{table}

After working out the direct product of $\Psi(q^3)$ and $\Psi(c\bar{c})$, all the quark configurations of $P_c$ and corresponding states are shown in Table \ref{Tab:pcconfig}, where $\phi$ and $\chi$ are the flavor and spin parts. 
The spatial wave functions are symmetric and are not specified here.
The last two configurations with the flavor singlet in Table \ref{Tab:pcconfig} are for the $q^3=uds$ light quark cluster and will not be considered in this study.

\begin{table*}[!ht]
\caption{Configurations of $P_c$ and possible quantum numbers.\label{Tab:pcconfig}}
\def\arraystretch{1.5}
\begin{ruledtabular}
\begin{tabular}{l c c c}
$P_c$ configurations & $\Psi(q^3)$ part \hspace{2.5cm} $\Psi(c \bar c)$ part & Isospin & Spin\\
\hline
$\Psi_{[111]_C[3]_F[3]_S [\chi_1]}$ &\makecell[l]{$\psi^C_{[111]} \phi_{[3]} \chi_{[3]} \hspace{1 cm} \otimes \hspace{1 cm} \psi^C_{[111]} \phi(c\bar c) \chi_{1}$} & $\frac{3}{2}$   & $\frac{1}{2}$, $\frac{3}{2}$, $\frac{5}{2}$\\
$\Psi_{[111]_C[3]_F[3]_S [\chi_0]}$&\makecell[l]{$\psi^C_{[111]} \phi_{[3]} \chi_{[3]} \hspace{1 cm} \otimes \hspace{1 cm} \psi^C_{[111]} \phi(c\bar c) \chi_{0}$} & $\frac{3}{2}$   & $\frac{1}{2}$\\
$\Psi_{[111]_C[21]_F[21]_S [\chi_1]}$&\makecell[l]{$\psi^C_{[111]}\phi_{[21]}\chi_{[21]}\hspace{0.76 cm} \otimes \hspace{1 cm} \psi^C_{[111]} \phi(c\bar c) \chi_{1}$} & $\frac{1}{2}$   & $\frac{1}{2}$, $\frac{3}{2}$\\
$\Psi_{[111]_C[21]_F[21]_S [\chi_0]}$ &\makecell[l]{$\psi^C_{[111]}\phi_{[21]}\chi_{[21]}\hspace{0.76 cm} \otimes \hspace{1 cm} \psi^C_{[111]} \phi(c\bar c) \chi_{0}$} & $\frac{1}{2}$   & $\frac{1}{2}$\\
\hline
$\Psi_{[21]_C[3]_F[21]_S [\chi_1]}$&\makecell[l]{$\psi^C_{[21]}\phi_{[3]} \chi_{[21]}\hspace{1.02 cm} \otimes \hspace{1 cm} \psi^C_{[21]} \phi(c\bar c) \chi_{1}$}& $\frac{3}{2}$   & $\frac{1}{2}$, $\frac{3}{2}$\\
$\Psi_{[21]_C[3]_F[21]_S [\chi_0]}$&\makecell[l]{$\psi^C_{[21]}\phi_{[3]} \chi_{[21]}\hspace{1.02 cm} \otimes \hspace{1 cm} \psi^C_{[21]} \phi(c\bar c) \chi_{0}$} & $\frac{3}{2}$   & $\frac{1}{2}$\\
$\Psi_{[21]_C[21]_F[3]_S [\chi_1]}$&\makecell[l]{$\psi^C_{[21]}\phi_{[21]}\chi_{[3]}\hspace{1.02 cm} \otimes \hspace{1 cm} \psi^C_{[21]} \phi(c\bar c) \chi_{1}$} & $\frac{1}{2}$   & $\frac{1}{2}$, $\frac{3}{2}$, $\frac{5}{2}$\\
$\Psi_{[21]_C[21]_F[3]_S [\chi_0]}$&\makecell[l]{$\psi^C_{[21]}\phi_{[21]}\chi_{[3]}\hspace{1.02 cm} \otimes \hspace{1 cm} \psi^C_{[21]} \phi(c\bar c) \chi_{0}$} & $\frac{1}{2}$   &  $\frac{3}{2}$\\
$\Psi_{[21]_C[21]_F[21]_S [\chi_1]}$&\makecell[l]{$\psi^C_{[21]}\phi_{[21]} \chi_{[21]}\hspace{0.90 cm} \otimes \hspace{1 cm}\psi^C_{[21]} \phi(c\bar c) \chi_{1}$} & $\frac{1}{2}$   & $\frac{1}{2}$, $\frac{3}{2}$\\
$\Psi_{[21]_C[21]_F[21]_S [\chi_0]}$&\makecell[l]{$\psi^C_{[21]}\phi_{[21]} \chi_{[21]}\hspace{0.9 cm} \otimes \hspace{1 cm} \psi^C_{[21]} \phi(c\bar c) \chi_{0}$} & $\frac{1}{2}$   & $\frac{1}{2}$\\
$\Psi_{[21]_C[111]_F[21]_S [\chi_1]}$&\makecell[l]{$\psi^C_{[21]}\phi_{[111]} \chi_{[21]}\hspace{0.76 cm} \otimes \hspace{1 cm} \psi^C_{[21]} \phi(c\bar c) \chi_{1}$}& 0   & $\frac{1}{2}$, $\frac{3}{2}$\\

$\Psi_{[21]_C[111]_F[21]_S [\chi_0]}$&\makecell[l]{$\psi^C_{[21]}\phi_{[111]} \chi_{[21]}\hspace{0.76 cm} \otimes \hspace{1 cm} \psi^C_{[21]} \phi(c\bar c) \chi_{0}$} & 0  & $\frac{1}{2}$\\
\end{tabular}
\end{ruledtabular}
\end{table*}

\section{$P_c$ partial decay widths} \label{sec3}
\begin{figure}[H]
\centering
\includegraphics[scale=0.4]{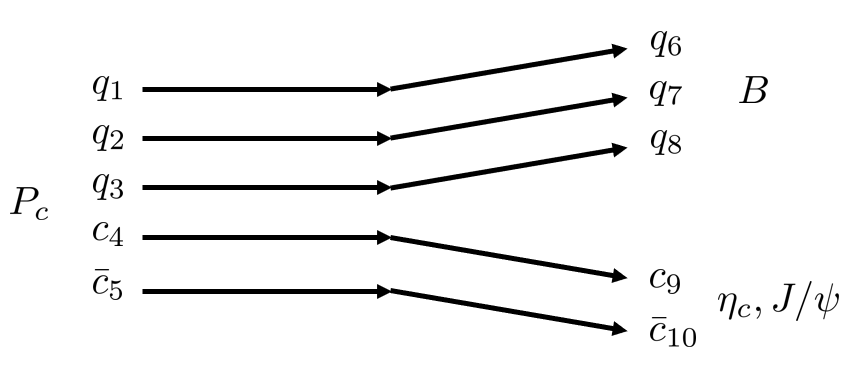}
\caption{Hidden charm decay process (direct process). \label{diagram1}}
\end{figure}

\begin{figure}[H]
\centering
\includegraphics[scale=0.36]{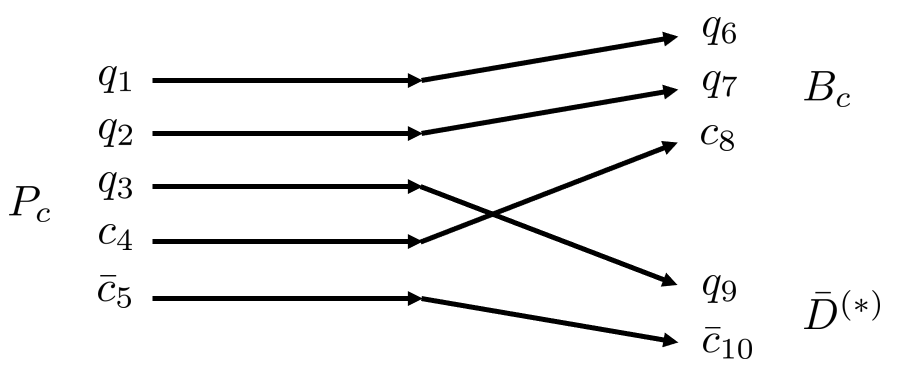}
\caption{Open charm decay process (cross process). \label{diagram2} }
\end{figure}

The $P_c$ resonances may decay through both the hidden-charm and open charm modes, as shown in Fig. \ref{diagram1} and Fig. \ref{diagram2}, respectively. 
The transition amplitudes are calculated from
	\begin{equation}
	\label{transition}
	T = \bra{\psi_{final}}\hat O\ket{\psi_{initial}},
	\end{equation}
where $\psi_{initial}$ represents the initial states of  hidden charm pentaquark listed in Table \ref{Tab:pcconfig} and $\psi_{final}$ represents the final states from all possible decay channels. The operator, $\hat O$, can be either $\hat O_d$  for the direct process or $\hat O_c$ for the cross process. These operators can be written in the form
\begin{align}
\label{delta1}
\hat O_{d}=& \lambda \; \delta^3(\vec{q}_1-\vec{q}_6)\delta^3(\vec{q}_2-\vec{q}_7)\delta^3(\vec{q}_3-\vec{q}_8)\delta^3(\vec{q}_4-\vec{q}_9) \nonumber\\&\delta^3(\vec{q}_5-\vec{q}_{10}),\\
\label{delta2}
\hat O_{c} =& \lambda \; \delta^3(\vec{q}_1-\vec{q}_6)\delta^3(\vec{q}_2-\vec{q}_7)\delta^3(\vec{q}_3-\vec{q}_9)\delta^3(\vec{q}_4-\vec{q}_8)\nonumber\\&\delta^3(\vec{q}_5-\vec{q}_{10}).
\end{align}
where $\lambda$ is a coupling constant which are assumed in this work the same for both direct and cross processes. 
	
Shown in Table \ref{Tab:spff} are the factors of  the spin-flavor-color transition amplitudes for the decays of 17 initial pentaquark states to 10 final states, $p\eta_c$, $\Delta\eta_c$, $p J/\psi$, $\Delta J/\psi$, $\Lambda_c \bar{D}$, $\Sigma_c \bar{D}$, $\Sigma^*_c \bar{D}$, $\Sigma_c \bar{D}^*$, $\Sigma^*_c \bar{D}^*$ and $\Lambda_c \bar{D}^*$. The pentaquark states in both the color singlet-singlet and color octet-octet configurations are classified by the isospins. 
\begin{table*}[!ht]
 \caption{Allowed spin-flavor-color transition amplitudes, $\bra{\psi_f^{SFC}} \hat O \ket{\psi_i^{SFC}}$.\label{Tab:spff}}
\def\arraystretch{1.5}
  \begin{ruledtabular}
   \begin{tabular}{c c c c c c c c c c c c}
    \multirow{1}{*}{$P_c$ Configuration} & \multirow{1}{*}{$J^P$} & $\Delta \eta_c$ &  $p \eta_c$ & $\Delta J/\psi$ & $p J/\psi$ & $\Sigma_c ^*\bar{D}$ & $\Sigma_c \bar{D}$ & $\Lambda_c \bar{D}$ &$\Sigma_c^*\bar{D}^*$ & $\Sigma_c\bar{D}^*$ & $\Lambda_c\bar{D}^*$  \\
      \hline
      & \multicolumn{11}{c}{Isospin 3/2}\\
      \hline
   $\Psi_{[111]_C[3]_F[3]_S [\chi_1]}$ & $\frac{5}{2}^-$ & & & 1 & & & & & $\frac{1}{3}$ & &  \\

   $\Psi_{[111]_C[3]_F[3]_S [\chi_1]}$ &  $\frac{3}{2}^-$ & & & 1 & & $\frac{\sqrt{\frac{5}{3}}}{6}$ & & &  $\frac{1}{8}$ & $\frac{\sqrt{5}}{9}$ & \\

   $\Psi_{[111]_C[3]_F[3]_S [\chi_0]}$ & $\frac{3}{2}^-$& & & 1 & & $\frac{1}{6}$ &  & & $\frac{\sqrt{\frac{5}{3}}}{6}$ & $-\frac{1}{3\sqrt{3}}$ &\\

   $\Psi_{[111]_C[3]_F[3]_S [\chi_1]}$ & $\frac{1}{2}^-$ & 1 & & & & & $\frac{\sqrt{\frac{2}{3}}}{3}$  & & $-\frac{1}{9}$ & $\frac{\sqrt{2}}{9}$ & \\

   \hline
   $\Psi_{[21]_C[3]_F[21]_S [\chi_1]}$ & $\frac{3}{2}^-$ & & & & & $\frac{2}{3\sqrt{3}}$ & & & $-\frac{2\sqrt{5}}{9}$ & $-\frac{2}{9}$ & \\

   $\Psi_{[21]_C[3]_F[21]_S [\chi_1]}$& $\frac{1}{2}^-$& & & & & & $\frac{1}{3\sqrt{3}}$ & &  $-\frac{2\sqrt{2}}{9}$ & $-\frac{5}{9}$ & \\

   $\Psi_{[21]_C[3]_F[21]_S [\chi_0]}$& $\frac{1}{2}^-$ & & & & & & $-\frac{1}{3}$ &  & $-\frac{2\sqrt{\frac{2}{3}}}{3}$ & $\frac{1}{3\sqrt{3}}$ &\\
      \hline
      & \multicolumn{11}{c}{Isospin 1/2}\\
      \hline
   $\Psi_{[111]_C[21]_F[21]_S [\chi_1]}$ & $\frac{3}{2}^-$ & & & & 1 & $-\frac{1}{3\sqrt{6}}$ &  & & $\frac{\sqrt{\frac{5}{2}}}{9}$ & $\frac{1}{9\sqrt{2}}$ & $\frac{1}{3\sqrt{2}}$ \\
   $\Psi_{[111]_C[21]_F[21]_S [\chi_1]}$ & $\frac{1}{2}^-$ & & & & 1 &  & $-\frac{1}{6\sqrt{6}}$&$\frac{1}{2\sqrt{6}}$ & $\frac{1}{9}$ & $\frac{5}{18\sqrt{2}}$ & $-\frac{1}{6\sqrt{2}}$ \\
   $\Psi_{[111]_C[21]_F[21]_S [\chi_0]}$ & $\frac{1}{2}^-$ & & 1 & & &  & $\frac{1}{6\sqrt{2}}$&$\frac{1}{6\sqrt{2}}$& $\frac{1}{3\sqrt{3}}$ & $-\frac{1}{6\sqrt{6}}$  & $\frac{1}{2\sqrt{6}}$ \\
   \hline
   $\Psi_{[21]_C[21]_F[3]_S [\chi_1]}$ & $\frac{5}{2}^-$ & & & & &  & &   & $-\frac{2}{3}$ &  &   \\
   $\Psi_{[21]_C[21]_F[3]_S [\chi_1]}$ & $\frac{3}{2}^-$ & & & &  & $-\frac{\sqrt{\frac{5}{3}}}{3}$ &  &  & $-\frac{1}{9}$ & $-\frac{2\sqrt{5}}{9}$ & \\
   $\Psi_{[21]_C[21]_F[3]_S [\chi_1]}$ & $\frac{1}{2}^-$ & & & &  & & $-\frac{2\sqrt{\frac{2}{3}}}{3}$ &   &$\frac{2}{9}$ &  $-\frac{2\sqrt{2}}{9}$ &\\
   $\Psi_{[21]_C[21]_F[3]_S [\chi_0]}$ & $\frac{3}{2}^-$ & & & & & $-\frac{1}{3}$ &  &  &  $-\frac{\sqrt{\frac{5}{3}}}{3}$ & $\frac{2}{3\sqrt{3}}$ & \\
   $\Psi_{[21]_C[21]_F[21]_S [\chi_1]}$ & $\frac{3}{2}^-$ & & & &  & $\frac{\sqrt{\frac{2}{3}}}{3}$ &  &  & $-\frac{\sqrt{10}}{9}$ & $-\frac{\sqrt{2}}{9}$ & $\frac{\sqrt{2}}{3}$ \\
   $\Psi_{[21]_C[21]_F[21]_S [\chi_1]}$ & $\frac{1}{2}^-$ & & & & & & $\frac{1}{3\sqrt{6}}$& $\frac{1}{\sqrt{6}}$ & $-\frac{2}{9}$ & $-\frac{5}{9\sqrt{2}}$  & $-\frac{1}{3\sqrt{2}}$ \\
   $\Psi_{[21]_C[21]_F[21]_S [\chi_0]}$  & $\frac{1}{2}^-$ & & & & &  & $-\frac{1}{3\sqrt{2}}$& $\frac{1}{3\sqrt{2}}$& $-\frac{2}{3\sqrt{3}}$ & $\frac{1}{3\sqrt{6}}$ & $-\frac{1}{\sqrt{6}}$\\
   \end{tabular}
  \end{ruledtabular}
 \end{table*}

With the transition amplitude defined in Eq.\ref{transition}, the decay width is evaluated in the non-relativistic approximation \cite{ajayut},
\begin{align}\label{transition2}
\Gamma_{P_c \rightarrow BM} = & \frac{2 \pi E_1 E_2}{ M_{P_c}} \frac{1}{2 S_i +1} \;f(m_B,m_M,q)\nonumber \\& \hspace{1.6cm} \sum_{m_i, m_f} |\bra{\psi_f^{SFC}} \hat O \ket{\psi_i^{SFC}}| ^2,
\end{align}
where $S_i$, $m_i$, and $M_{P_c}$ are the spin, spin projection quantum number, and mass of the initial pentaquark states, respectively. $E_1$ and $E_2$ are the energies of the baryon and meson in the final states, $\bra{\psi_f^{SFC}} \hat O \ket{\psi_i^{SFC}}$ is the spin-flavor-color factors of the transition amplitudes listed in Table \ref{Tab:spff}. The summation is over the spins of the initial and final states. $f(m_B,m_M,q)$ is a kinematical phase-space factor depending on the masses of the baryon and meson in the final states as well as the spatial wave functions of all particles. To reduce the model dependence, we apply the phenomenological form for the phase space factor, 
	\begin{equation}\label{phasefactor}
	f(m_B,m_M,q)= q \; \text{exp} \{-1.2 \;\text{GeV}^{-1}\;(s-s_{0})^{1/2}\},
	\end{equation}
where $s_{0}=(m_{B}+m_{M})^{2}$, $\sqrt{s}=(m_{B}^2+q^2)^{1/2}+(m_{M}^2+q^2)^{1/2} $, and $q$ is the final momentum at the rest frame of the pentaquark.
The kinematical phase-space factor in Eq. (\ref{phasefactor}) has been fitted to the cross section of various $p\bar p$ annihilation channels \cite{Vandermeulen1988} and applied successfully to other works \cite{GUTSCHE1997311,PhysRevC.59.630,Sorakrai2016}. The numerical values of the kinematical phase-space factor are shown in Appendix \ref{App:phasespacefactor} for the masses of $P_c(4312)$ $P_c(4440)$ and $P_c(4457)$ resonances and various decay channels. Some decay channels of $P_c$ states are vanished by missing of spin-flavor-color transition amplitude in Table \ref{Tab:spff} and phase space factor in Table \ref{Tab:phasespace}. It is seen that $P_c(4312)$ in the configuration $\psi_{[111]_C[21]_F[21]_S[\chi_1]}$ with spin $\frac{3}{2}$ has two decay channels, $p J/\psi$ and $ \Lambda_c \bar{D}^*$ while $P_c(4440)$ and $P_c(4457)$ with the same configuration have one more open charm decay channel, $\Sigma^*_c\bar{D}$. For the configuration $\psi_{[111]_C[21]_F[21]_S[\chi_1]}$ and spin $\frac{1}{2}$, $P_c(4312)$ has $p J/\psi$, $\Lambda_c\bar{D}$ and $\Lambda_c\bar{D}^*$ decay channels while $P_c(4440)$ and $P_c(4457)$ get one more open channel, $\Sigma_c\bar{D}$.

The partial decay widths of all available pentaquark states are calculated in Eq. \ref{transition2}. Listed in Table \ref{Pc4457normalized} are the results for only the $I=\frac{1}{2}$ pentaquark states, 
where the partial decay widths are normalized to the $\psi_{[111]_C[21]_F[21]_S[\chi_1]}\rightarrow p J/\psi$ process. In the calculation we let the pentaquark states take the mass of $P_c(4457)$. The results for the channels allowed by the energy conservation are similar when the masses of $P_c(4312)$, $P_c(4440)$ are applied for the initial pentaquark states.
In the calculation, we have considered only the $S$-wave decay since the mass thresholds for all the final states are very close to the mass of the $P_c$ resonances. Among the 17 compact pentaquark states, the state of spin $5/2$ in the $\psi_{[21]_C[21]_F[3]_S[\chi_1]}$ configuration has no any allowed decay channel. 

\begin{table}[H]
\caption{Partial decay widths of $I = \frac{1}{2}$ pentaquark states of 4457 MeV, normalized to the $\psi_{[111]_C[21]_F[21]_S[\chi_1]}\rightarrow p J/\psi$ width.\label{Pc4457normalized}}
\def\arraystretch{1.5}
\begin{ruledtabular}
	\begin{tabular}{m{0.3cm} m{2.7cm} c c c c c c}
	
	 $J^p$ & $P_c$ configuration  & $p J/\psi$ & $p \eta_c$ & $\Sigma_c^* \bar{D}$ & $\Sigma_c \bar{D}$ & $\Lambda_c  \bar{D}$ & $\Lambda_c  \bar{D}^*$ \\
\hline\hline
	 $\frac{3}{2}^-$  & $\psi_{[111]_C[21]_F[21]_S[\chi_1]}$ & \makecell[c]{ 1 } & & \makecell[l]{ 0.04 } & & & \makecell[l]{ 0.11}  \\
	  & $\psi_{[21]_C[21]_F[3]_S[\chi_1]}$ & & & \makecell[c]{ 0.40} & & &  \\
	  & $\psi_{[21]_C[21]_F[3]_S[\chi_0]}$  & & & \makecell[c]{ 0.24} & & & \\
	  & $\psi_{[21]_C[21]_F[21]_S[\chi_1]}$  & & & \makecell[c]{ 0.16} & & & \makecell[c]{0.46} \\
	\hline
	 $\frac{1}{2}^-$  & $\psi_{[111]_C[21]_F[21]_S[\chi_1]}$  & \makecell[c]{ 1 } & & & \makecell[c]{ 0.01 } & \makecell[c]{ 0.07} & \makecell[c]{ 0.03}  \\
	  & $\psi_{[111]_C[21]_F[21]_S[\chi_0]}$  & & \makecell[c]{ 0.90 } & & \makecell[c]{0.03} & \makecell[c]{ 0.02} & \makecell[l]{ 0.09} \\
	  & $\psi_{[21]_C[21]_F[3]_S[\chi_1]}$ & & & & \makecell[c]{ 0.62 } & &  \\
	  & $\psi_{[21]_C[21]_F[21]_S[\chi_1]}$  & & & & \makecell[c]{0.04} & \makecell[c]{ 0.28} & \makecell[c]{ 0.11}  \\
	  & $\psi_{[21]_C[21]_F[21]_S[\chi_0]}$  & & &  & \makecell[c]{ 0.12} & \makecell[c]{ 0.09} & \makecell[c]{ 0.34}  \\		
	
\end{tabular}
\end{ruledtabular}
\end{table}

\section{DISCUSSION}\label{sec4}

We have calculated the partial decay widths for all the 17 compact pentaquark states since the quantum numbers of the three LHCb $P_c$ resonances have not been determined yet. The results in Table \ref{Pc4457normalized} for the nine $I=1/2$ pentaquark states show that the $pJ/\psi$ channel is open for only two states with the configuration $\psi_{[111]_C[21]_F[21]_S[\chi_1]}$ and spin $\frac{3}{2}$ and $\frac{1}{2}$.	
%If the four $I=\frac{1}{2}$ and $J=\frac{3}{2}$ states could linearly mix, there could be four physical mixing states. The same goes with the five $I=\frac{1}{2}$ and $J=\frac{1}{2}$ states as shown in Table \ref{Pc4457normalized}. Therefore, there could possibly be nine compact charmonium-like pentaquark states, which may decay into $pJ/\psi$. 
If there is no mixing among the $I=\frac{1}{2}$ and $J=\frac{3}{2}$ states as well as among the $I=\frac{1}{2}$ and $J=\frac{1}{2}$ states, there are only two compact charmonium-like pentaquark states which decay through the $pJ/\psi$ channel. Therefore, one may accommodate only two of the three observed $P_c$ in the compact pentaquark picture. Our results show that the $pJ/\psi$ partial decay widths of $P_c(4312)$, $P_c(4440)$ and $P_c(4457)$ in the configuration $\psi_{[111]_C[21]_F[21]_S[\chi_1]}$ are almost the same due to the mass closeness, that is 
\begin{eqnarray}
\frac{P_c(4312)\rightarrow pJ/\psi}{P_c(4457)\rightarrow pJ/\psi} &=& 1.16 \nonumber \\
\frac{ P_c(4440)\rightarrow pJ/\psi}{P_c(4457)\rightarrow pJ/\psi} &=& 1.02
\end{eqnarray}
Thus, this work indicates that $P_c(4440)$ may not be a compact pentaquark since its decay width is much larger than the other observed $P_c$, and that one may assign $J=\frac{1}{2}$ to the $P_c(4312)$ state and $J=\frac{3}{2}$ to the $P_c(4457)$ state in the compact pentaquark picture.
	
%	Based on the decay ratios in the same table, we suggest charmonium-like pentaquarks to be searched in other channels too, especially in the $p\eta_c$ channel. For the four possible open charm decay channels, they may probably be found in a future experiment even though the partial width ratios are small when they are normalized by $P_c(4457)$ $\rightarrow$ $pJ/\psi$.
	
The branching ratios of the dominant channels are the key to reveal the nature of these charmonium-like pentaquark states. In addition to the $p J/\psi$ decay channel, the remaining possible decay channels, particularly the $p\eta_c$ channel should be targeted to search for $P_c$ states and determine their quantum numbers in the future experiments.

\section*{ACKNOWLEDGEMENTS}
\indent
This work is supported by Suranaree University of Technology
(SUT), National Cheng Kung University(NCKU) and Development and Promotion of Science and Technology Talents Project (DPST).

\appendix

\section{The explicit color wave function} \label{App:colorwave}
The explicit color singlet and octet wave functions of the $q^3$ and $c\bar c$ clusters are listed in Table \ref{Tab:Cwbpp}, which are employed to evaluate the spin-flavor-color transition amplitude of pentaquark states.

\begin{table*}[ht]
	\caption{Color wave functions of baryon and meson clusters for pentaquarks.
	\label{Tab:Cwbpp}}
	\renewcommand{\arraystretch}{1.5}
	\centering
	\begin{ruledtabular}
	\begin{tabular}{c l l l}
	 & \makecell[c]{$q^3$}  & & \makecell[c]{$Q\bar{Q}$} \\
	\hline
	\makecell{Singlet} & \makecell[l]{$\frac{1}{\sqrt{6}}( RGB-RBG+GBR-GRB$\\$+BRG-BGR) $}& & \makecell[l]{$\frac{1}{\sqrt{3}}(R\bar{R}+G\bar{G}+B\bar{B})$}  \\
	\hline\hline
	 & \makecell[c]{$q^3 \rho$-type}  & \makecell[c]{$q^3 \lambda$-type} & \makecell[c]{$Q\bar{Q}$}\\
	\hline
	\makecell{Octet} & & & \\
	 1 &$\frac{1}{\sqrt{2}}(RGR-GRR)$  & $\frac{1}{\sqrt{6}}(RRG-RGR-GRR)$ & $B\bar{R}$\\
	 2 &$\frac{1}{\sqrt{2}}(RGG-GRG)$  & $\frac{1}{\sqrt{6}}(RGG-GRG-2GGR)$ & $B\bar{G}$\\
	 3 &$\frac{1}{\sqrt{2}}(RBR-BRR)$  & $\frac{1}{\sqrt{6}}(2RRB-RBR-BRR)$ & $-G\bar{R}$\\
	 4 & \makecell[l]{$\frac{1}{2}(RBG+GBR-BRG-BGR)$}  & \makecell[l]{$\frac{1}{\sqrt{12}}(2RGB+2GRB-GBR-RBG$\\$-BRG-BGR)$} & $\frac{1}{\sqrt{2}}(R\bar{R}-G\bar{G})$\\
	 5 &$\frac{1}{\sqrt{2}}(GBG-BGG)$  & $\frac{1}{\sqrt{6}}(2GGB-GBG-BGG)$ & $R\bar{G}$\\
	 6 & \makecell[l]{$\frac{1}{\sqrt{12}}(2RGB-2GRB-GBR+RBG$\\$-BRG+BGR)$} & \makecell[l]{$\frac{1}{2}(RBG-GBR+BRG-BGR)$} & \makecell[l]{$\frac{1}{\sqrt{6}}(2B\bar{B}-R\bar{R}-G\bar{G})$}\\
	 7 &$\frac{1}{\sqrt{2}}(RBB-BRB)$ & $\frac{1}{\sqrt{6}}(RBB+BRB-2BBR)$ & $-G\bar{B}$ \\
	 8 &$\frac{1}{\sqrt{2}}(GBB-BGB)$ & $\frac{1}{\sqrt{6}}(GBB+BGB-2BBG)$ & $R\bar{B}$ \\
	\end{tabular}
	\end{ruledtabular}
\end{table*}

\section{Phase space factor with the possible decay channels} \label{App:phasespacefactor}
Listed in Table \ref{Tab:phasespace} are the numerical values of the kinematical phase-space factor in Eq. \eqref{phasefactor}. Some decay channels are not available (N/A) because of the energy conservation.

\begin{table*}[ht]
	\caption{Phase space factor, $f(m_B,m_M,q)$, for $P_c(4312)$, $P_c(4440)$ and $P_c(4457)$ with possible decay channels.\label{Tab:phasespace}}
	\renewcommand{\arraystretch}{1.5}
		\begin{center}
		\begin{ruledtabular}
			\begin{tabular}{ccc c c }
			
				\multirow{2}{*}{Final particles} & Total mass of  & \multirow{2}{*}{$P_c(4312)$} & \multirow{2}{*}{$P_c(4440)$} & \multirow{2}{*}{$P_c(4457)$}\\
				&final particles (GeV)&&&\\
			\hline
			$p \eta_c$ & 3.922 & 0.08056 & 0.07038 & 0.06911\\
			$\Delta \eta_c$ & 4.214 & 0.12653 & 0.11442 & 0.11221\\
			$p J/\psi$ & 4.035 & 0.08929 & 0.07827 & 0.07686\\
			$\Delta J/\psi$ & 4.327 & N/A & 0.12631 & 0.12516\\
			$\Lambda_c \bar{D}$ & 4.157 & 0.15350 & 0.13097 & 0.12797\\
			$\Sigma_c \bar{D}$ & 4.325 & N/A & 0.16237 & 0.16014\\
			$\Sigma_c^* \bar{D}$ & 4.390 & N/A & 0.16176 & 0.16517\\
			$\Lambda_c \bar{D}^*$ & 4.297 & 0.12507 & 0.16132 & 0.15847\\
			$\Sigma_c \bar{D}^*$ & 4.465 & N/A & N/A & N/A\\
			$\Sigma_c^* \bar{D}^*$ & 4.530 & N/A & N/A & N/A\\
			
			\end{tabular}
			\end{ruledtabular}
		\end{center}
	\end{table*}
	
\newpage
\bibliographystyle{unsrt}
%\bibliography{/home/thiago/bibtex/articles,/home/thiago/bibtex/books}

\end{document}